\documentclass[twoside,twocolumn,9pt]{revtex4}
\usepackage{amsmath}
\usepackage{amsfonts}
\usepackage{amssymb}
\usepackage{mathrsfs}
\usepackage{graphicx}
\begin{document}
\title{Pathways connecting two opposed bilayers with a fusion pore: 
A molecularly-informed phase field approach}

\author{Yucen Han$^{1}$}
\author{Zirui Xu$^{2}$}
\author{An-Chang Shi$^{3}$}
\author{Lei Zhang$^{4}$}
\affiliation{$^1$ Beijing International Center for Mathematical Research, Peking University, Beijing 100871, China.\\
$^2$Department of Applied Physics and Applied Mathematics, Columbia University, New York, NY 10027, USA.\\
$^3$Department of Physics and Astronomy, McMaster University, Hamilton, Canada L8S 4M1.\\
$^4$Beijing International Center for Mathematical Research, Center for Quantitative Biology, Peking University, Beijing 100871, China.}

\begin{abstract}
A phase field model with two phase fields, representing the concentration and the head-tail separation of amphiphilic molecules, respectively, 
has been constructed using an extension of the Ohta-Kawasaki model ({\it Macromolecules} {\bf 19}, 2621-2632 (1986)). It is shown that this molecularly-informed  
phase field model is capable of producing various self-assembled amphiphilic aggregates, such as bilayers, vesicles and micelles. 
Furthermore, pathways connecting two opposed bilayers with a fusion pore are obtained by using a combination of the phase field model and the string method.
Multiple fusion pathways, including a classical pathway and a leaky pathway, have been obtained depending on the initial separation of the two bilayers.
    The study shed light to the understanding of membrane fusion pathways and, more importantly, laid a foundation for further investigation of more 
complex membrane morphologies and transitions.
\end{abstract}

\maketitle
\section{Introduction}

Membrane fusion is a structural transformation in which two initially separated lipid bilayers merge their hydrophobic cores, 
resulting in one interconnected structure. The fusion of two opposed lipid bilayers is a fundamental step in many important biological 
processes ranging from endocytosis and exocytosis, synaptic release, and viral infection to fertilization~\cite{chernomordik1995lipids,monck1996fusion,zimmerberg1999membrane,jahn2002membrane,tamm2003membrane,blumenthal2003membrane}.
The fusion transformation rate and structure depend sensitively on the transition pathway connecting different intermediate states of the system. 
Therefore, understanding the structure and energetics of fusion pathways is very important and it has attracted much attention.
An intuitively attractive pathway, the so-called classical pathway, connecting two opposed bilayers to a fusion pore was proposed by Kozlov and Markin~\cite{kozlov1983possible,markin1983primary}.
It is assumed that all the morphologies or structures along this pathway are rotationally symmetric. As the first step of the fusion transition, 
a highly curved intermediate structure, the so-called stalk~\cite{kozlovsky2002stalk, kozlovsky2004stalk}, is formed by connecting two opposed \textit{cis} monolayers. 
In the second step, the displacement of the two inner \textit{cis} leaves and  two outer \textit{trans} leaves results in the formation of a structure termed hemifusion diaphragm\cite{kozlovsky2002lipid,katsov2004field}. 
Finally, the rupture of this hemifusion diaphragm creates a fusion pore\cite{monck1996fusion,ryham2013teardrop}.
Besides the classical pathway, other fusion pathways have been proposed. One non-classical pathway is the so-called leaky 
pathway where a stalk-hole complex configuration\cite{katsov2006field,muller2003new} is formed.
Some evidence of the formation of the leaky fusion pathway has been reported in some experimental  
studies~\cite{shangguan1996influenza,bonnafous2000membrane,dunina2000hemagglutinin,haque2002influence,frolov2003membrane}, 
revealing that there is a path connecting the interior of the cells to the outside. 
Another evidence of the leaky fusion pathways is that mixing of the lipids in the \textit{cis} leaves and \textit{trans} leaves has been reported~\cite{lee1997evolution,evans2002kinetics}.

In the literature, various membrane models have been developed to study the structure of membranes and the process of membrane fusion. 
Atomistic models are able to provide details of the molecular structure. However, molecular dynamics (MD) simulations of atomistic model of lipid bilayers~\cite{damodaran1992structure} 
could only extend to systems of the order of a few nm over a time span about 0.1ns. The huge disparity between the capability of MD simulations and the length and time 
scale of membrane fusion made it virtually impossible to simulate the fusion pathways using atomistic model. In order to overcome this barrier, 
coarse-grained models, such as the Minimal model~\cite{carmesin1988bond,goetz1998computer} and the Systematic coarse-graining model~\cite{yelash2006well}, 
have been introduced. These models reduce the number of degrees of freedom, resulting in a significant computational speed-up at the cost of atomistic details.
At an even coarser scale, continuum models, such as the Helfrich elasticity model~\cite{kozlovsky2002stalk}, have been developed for the study of membrane structures and morphologies.
It is desirable to develop theoretical models at the mesoscopic scale that could connect the molecular and continuum models. 

One possible framework to develop a mesoscopic model for membrane fusion is the phase field model, which is a popular approach for the study of moving interface problems~\cite{van1979thermodynamic}.
It is natural to apply this framework to the study of membrane fusion because the membrane could be regarded as a boundary separating the \textit{trans} and \textit{cis} sides of the cell. 

Within the framework of the phase field model, the physical state of a particular system is described by a continuous phase field which takes different constant values in different phases. 
The different phases are further connected via smooth interfaces that are in turn determined by certain energy functional of the phase field. 
Over the years, various phase field models have been developed and applied to a broad-range of problems,
such as solidification, solid-state phase transition, coarsening and grain growth, {\it etc}~\cite{kobayashi1993modeling,li2006temperature,nestler1999multiphase}.
The application of phase-field method in study of vesicle membrane is first proposed by Du {\it et al.}~\cite{du2004phase}, 
initially focusing on the shape transformation of single component vesicle. 
Later applications of phase field model to membranes include the structure of a multi-component vesicle~\cite{wang2008modelling}, 
a vesicle interacting with a surrounding fluid~\cite{du2009energetic}, and the adhesion of a vesicle to a substrate~\cite{zhang2009phase}.
These studies provide interesting understanding of and insight into the morphological change of vesicles. 
However, in these phase field studies the lipid membranes are treated as a structureless elastic surface. 
As such, important structural information about the bilayer nature of the membrane is missing in these simple phase field models.

In this paper, we fill this gap by proposing a phase field model that contains molecular information of the amphiphiles. 
Specifically, we introduce two phase fields, representing the concentration distribution and head-tail separation of the amphiphilic molecules, respectively.   
The energy (or free energy) functional governing the formation and structure of the bilayers is derived based on the Ohta-Kawasaki model,
which is a free energy functional of a specific amphiphilic molecules, i.e., diblock copolymers. 
The two phase fields describe the macroscopic phase separation between the amphiphiles and solvent and microscopic phase separation 
between the hydrophilic heads and hydrophobic tails of the amphiphiles, respectively. 
The energy functional of the proposed phase field model is in the form of a generalized Landau free energy functional containing a number of parameters. 
The molecular interactions between solvent, amphiphile head and amphiphile tail are encoded in these parameters. 
Within this molecularly-informed phase field model, the bilayer structure is produced naturally from minimization of the free energy.  
Therefore, the resulting fusion pathways automatically contain the structural detail of the bilayers.
The advantage of this model is its ability to provide a more faithful description of membrane structures and transition pathways,
in contrast to the previously proposed simple phase field model with one phase field. 
As a first step to validate the model, we have obtained numerically the bilayer, liposome and micelle, as equilibria configurations of the model.
Furthermore, we also obtained at least four solutions (bilayers, a stalk, a diaphragm and fusion pore) that are solutions of the Euler-Lagrange equation of the free energy functional.
These solutions correspond to the intermediate states along the classical fusion pathway.
Previous studies of membrane fusion mainly focused on the transition from two apposing bilayers to the stalk~\cite{smirnova2019thermodynamically,muller2012transition}
rather than the entire transition pathway from two opposing bilayers to stalk, from stalk to hemifusion diaphragm and from hemifusion diaphragm to fusion pore. 
In the current study, we are able to obtain the full fusion pathways using the new phase field model.
In what follows, we present two complete membrane fusion pathways. Specifically, when the two opposed bilayers are in close proximity, 
we find a symmetric classical  fusion pathway. However, when the two opposed bilayers are separated by a gap, 
we find an asymmetric leaky fusion pathway which passes through a stalk-hole complex configuration. 

The rest of the paper is organized as follows. In Section II, we present the derivation and explanation of the phase field model in the form of a generalized Landau free energy functional. 
In Section III, we describe the numerical method used in the current work to compute the free energy minimizers and the transition pathways. 
In Section IV, we give the main results on the fusion pathways. Finally, Section V contains some concluding remarks.

\section{Model Development}

Bilayer membranes are self-assembled from lipid molecules in water. As such, the development of a phase field model for the
study of bilayer membranes must reflects the fact that lipids are amphiphilic molecules composed of a hydrophilic head and hydrophobic tails. 
As a generic model, we assume that the system is composed of amphiphilic molecules with a hydrophilic head A and hydrophobic tail B and solvents C. 
If a detailed model of the molecules is given, the property of the system could be studied by using standard tools of statistical mechanics such as 
mean-field theory and simulations. On the other hand, our purpose is to develop a phase field model containing information about the essential
feature of different molecular species. Our derivation of the phase field model is motivated by the work of Ohta and Kawasaki, who have derived
a Landau free energy functional of diblock copolymers, which could be regarded as a generic model of amphiphilic molecules. Specifically, we use
the functional form of the free energy expansion developed for a system composed of AB diblock copolymers and C homopolymers~\cite{ohta1995dynamics}. Treating the diblock 
copolymers as an analogue of a lipid molecule and the homopolymers as solvents, we can write down a generic Landau free energy for the system.
In the original form of the Ohta-Kawasaki (OK) model \cite{ohta1986equilibrium}, the parameters of the free energy expansion are derived from the molecular
model. In our approach, we will adopt the functional form of the free energy while treat the parameters of the model as phenomenological parameters, 
as commonly done in the development of phase field models.

Specifically, we assume that the state of the system is described by two independent phase fields or order parameters, $\eta$ and $\phi$.
Here the phase field $\eta$ represents the concentration of the amphiphilies (AB copolymers), thus  $\eta$  is an order parameter or phase field describing the 
macroscopic smacroscopic phase separationeparation of the amphiphiles and the solvents. On the other hand, the phase field $\phi$ represents the difference between the local volume 
fraction of the head A and tail B, so it describes the microscopic phase separation between the heads and tails.
Following the derivation of Ohta and Kawasaki and assuming that our system is incompressible, the free energy of the mixture of amphiphiles and solvents
can be written as a sum of short-range and long-range contributions 
\cite{ohta1995dynamics},
\begin{equation}
F\{\eta,\phi\} = F_S\{\eta,\phi\} + F_L\{\eta,\phi\},
    \label{F}
\end{equation}
where the short range contribution $F_S$ to the free energy is given by,
\begin{equation}
F_S\{\eta,\phi\}=\int\rm{d}\mathbf{r}\Big[\frac{c_1}{2}\vert\nabla\eta\vert^2+\frac{c_2}{2}\vert\nabla\phi\vert^2+W(\eta,\phi)\Big],
\end{equation}
where the coefficients $c_1$ and $c_2$ are parameters governing the thickness of macroscopic phase separation interface and microscopic phase separation interface, respectively.
In the expression of $F_S$, the local contribution is specified as a polynomial of the phase fields,
\begin{equation}
    \begin{aligned}
        W(\eta,\phi) &= \frac{(\eta^2\!-\!1)^2}{4}+\frac{(\phi^2\!-\!1)^2}{4}\\
        &+b_1\eta\phi-\frac{b_2}{2}\eta\phi^2-\frac{b_3}{2}\phi\eta^2+\frac{b_4}{2}\eta^2\phi^2.
    \end{aligned}
    \label{W}
\end{equation}
The first two terms in Eq. \ref{W} exhibit double-well potential for $\eta$ and $\phi$, respectively, and the rest terms in Eq. \ref{W} describe the coupling between the amphiphiles (AB copolymers) and the solvents (C). The coefficients $b_1$, $b_2$ and $b_4$ are positive constants. 
In the original Ohta-Kawasaki model developed for mixtures of diblock copolymers and homopolymers, these parameters are related to 
the molecular parameters~\cite{avalos2016frustrated, ito1998domain, ohta1995dynamics} and could be derived in principle by the generalized method given in~\cite{ohta1986equilibrium}.
However, in the current work we will treat these parameters as phenomenological ones. We will adjust these parameters such that the potential $W(\eta,\phi)$ has
a triple-well potential structure with three distinct minima corresponding to the phases of head A, tail B and solvent C, respectively. Furthermore, the parameters will be chosen such that 
the head A-segments are localized at the interface between solvents and amphiphiles.

The fact that the head and tail of the amphiphilic molecules are linked together leads to the long-range contribution $F_L$ to the free energy. Following the
Ohta-Kawasaki model, the long-range contribution can be written in the form,
\begin{equation}
\begin{aligned}
    F_L\{\eta,\phi\}&=\int\!\!\rm{d}\mathbf{r}\!\int\!\!\rm{d}\mathbf{r'}\;G(\mathbf{r},\mathbf{r'})\big[\frac{\alpha}{2}\delta\phi(\mathbf{r})\delta\phi(\mathbf{r'})\\
    &+\beta\delta\phi(\mathbf{r})\delta\eta(\mathbf{r'})
+\frac{\gamma}{2}\delta\eta(\mathbf{r})\delta\eta(\mathbf{r'})\big],
\end{aligned}
\end{equation}
where the Green's function ${\rm G}(\mathbf{r},\mathbf{r'})$ is defined by the relation $-\nabla^2{\rm G}(\mathbf{r},\mathbf{r'})=\delta(\mathbf{r}\!-\!\mathbf{r'})$, 
the field variables $\delta \eta = \eta-\bar{\eta}, \delta \phi = \phi - \bar{\phi}$ with $\bar{\eta}$ and $\bar{\phi}$ representing the spatial average of $\eta$ and $\phi$, respectively.
%
Again the coefficients $\alpha$, $\beta$ and $\gamma$ are related to the molecular parameters of the model system, For the case of mixtures of
diblock copolymers and homopolymers, these coefficients are specified by the molecular weights of the different blocks. In the current work, we take
these coefficients as phenomenological parameters. For simplicity, we assume $\beta = \gamma = 0$ and take $\alpha$ as a positive coefficient.
When $\alpha = 0$, there is no linkage between the head and tail so the system would phase separate macroscopically. 
For the cases with $\alpha > 0$, the long range interaction originated from the head-tail linkage, would lead to microscopic phase separation, 
which is the key to producing the various membrane configuration such as bilayer, stalk, etc. 

\begin{figure}
        \includegraphics[width=1\columnwidth]{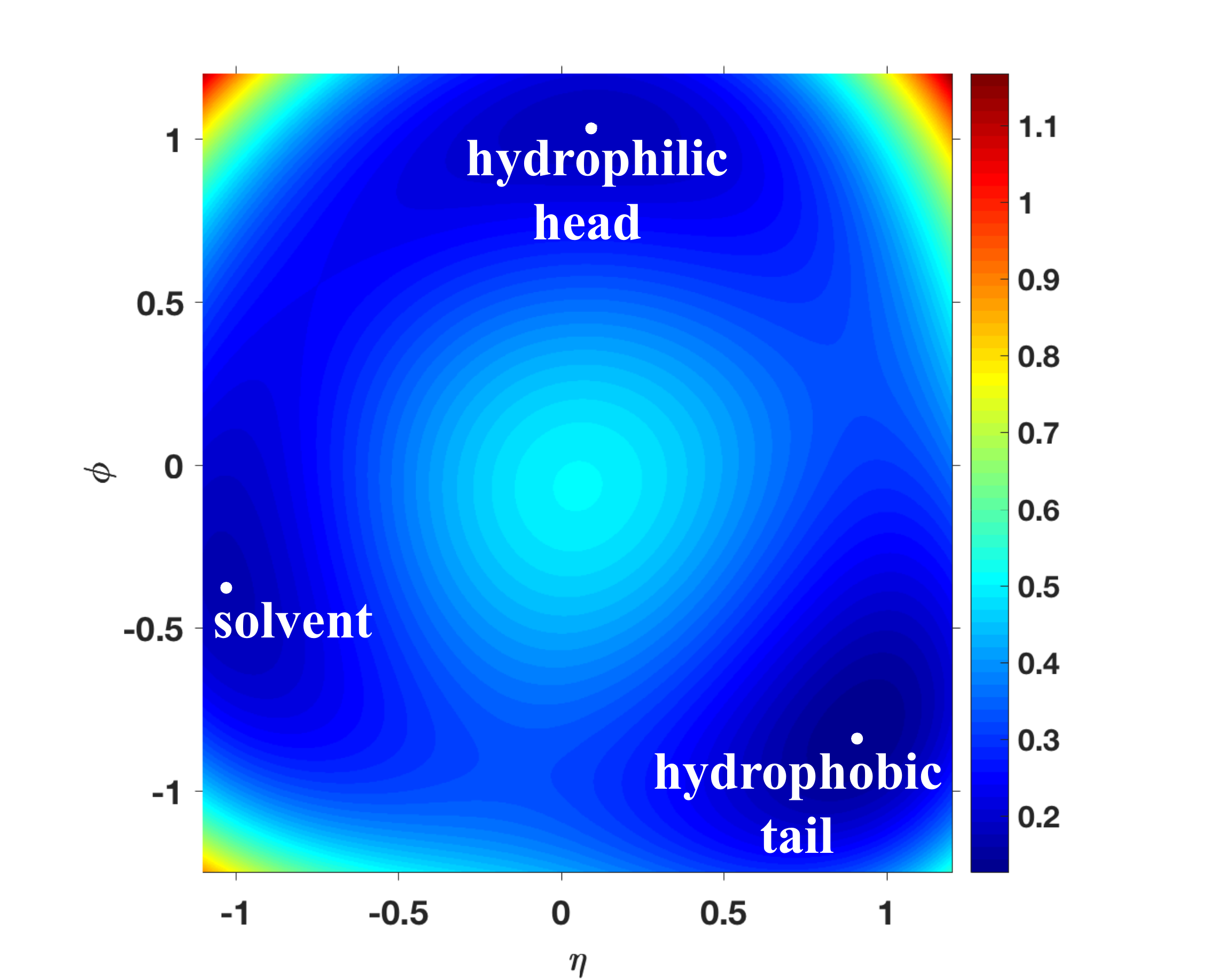}
    \caption{Contour plot of $W(\eta,\phi)$ given in Eq. \ref{W}. Three minima represent hydrophilic head, hydrophobic tail and solvent, respectively.}
    \label{local}
\end{figure}

The values of the coefficients specifying the free energy are chosen according to the physical behaviour of the system. 
After extensive numerical tests, we decided to choose $b_1 =0.04$, $b_2=0.25$, $b_3=0.425$ and $b_4=0.85$, such that the bulk free energy $W(\eta,\phi)$ has three
minima as shown in Fig. \ref{local}.  These minima are located at $(\eta,\phi)=(0.915,-0.831)$ for hydrophobic tail, $(\eta,\phi)=(0.106,1.038)$ for hydrophilic head and $(\eta,\phi)=(-1.029,-0.378)$ for solvent. 
We further assume $\alpha = 0.008$ and $c_1 = 0.6, c_2 = 1$. Our numerical solutions showed that this choice of the parameters produced reasonable solutions corresponding 
to self-assembled bilayer membranes.

\section{Numerical Method}

Within the framework of the phase field model, the task of studying stable and metastable structures of the model system is to obtain solutions of
the Euler-Lagrange equation of the free energy functional, that is, to obtain solutions minimizing the free energy. For the free energy functional given
above, the Euler-Lagrange equations are specified by $$\frac{\delta F}{\delta \eta}=\frac{\delta F}{\delta \phi}=0.$$ Instead of solving this set of 
nonlinear and non-local equations directly, it is more efficient to obtain the solutions using a dynamic approach. Because the phase fields of our model
are conserved order parameters satisfying the constraints $\int_{\Omega} (\eta-\bar{\eta})\rm{d}\mathbf{r} = 0$ and $\int_{\Omega} (\phi-\bar{\phi}) \rm{d}\mathbf{r} = 0$, 
the most convenient dynamic model is the Cahn-Hilliard model designed for conserved order parameters. 
Accordingly, the stable and metastable structures of our phase field model are obtained as the steady-state solutions of two coupled dynamic equations,
\begin{equation}
    \begin{aligned}
        \frac{\partial \eta}{\partial t} &= L_1\nabla^2\frac{\delta F}{\delta \eta} = L_1(\nabla^2f_1-\nabla^2\eta\\
        &+b_1\nabla^2\phi-c_1\nabla^4\eta
        -\beta\delta\phi-\gamma\delta\eta),\\
        \frac{\partial \phi}{\partial t} &= L_2\nabla^2\frac{\delta F}{\delta \phi} = L_2(\nabla^2f_2-\nabla^2\phi\\
        &+b_1\nabla^2\eta-c_2\nabla^4\phi
        -\alpha\delta\phi-\beta\delta\eta),
    \end{aligned}
    \label{Cahn-Hilliard}
\end{equation}
where $L_1$ and $L_2$ are mobility coefficients that will be set as $L_1 = L_2 = 1$ in our numerical calculations.
Furthermore, for convenience we have defined two nonlinear terms specified by,
\begin{equation*}
\left\{
\begin{aligned}
    f_1\{\eta,\phi\}&=\eta^3-\frac{b_2}{2}\phi^2-b_3\phi\eta+b_4\eta\phi^2,\\
f_2\{\eta,\phi\}&=\phi^3-b_2\eta\phi-\frac{b_3}{2}\eta^2+b_4\eta^2\phi.
\end{aligned}
\right.
\end{equation*}
Starting from a given initial configurations, the phase fields will be evolved according to the the dynamic equations of Eq.\ref{Cahn-Hilliard}. The steady-states of the
dynamic equations are then taken as solutions, corresponding to different stable or metastable structures, of the phase field model.

The dynamic phase field model specified above is a generic model valid in any spatial dimensions. However, the results reported in the paper were obtained by carrying out numerical calculations of the system in two-dimension. 
This choice was made because the main focus of the current paper is to construct the phase field model based on molecular information and to demonstrate the validity of the model
using simple examples. 
Specifically, our simulations were carried out in a two-dimensional domain of the size $\Omega = \left[0,128\right]\times\left[0,256\right]$. The computational domain is discretized 
using a 128$\times$256 square lattice with periodic boundary conditions. We use a first-order semi-implicit scheme in time with $\Delta t = 0.2$. 
The discrete dynamic equations are given by,
\begin{equation*}
\left\{
\begin{aligned}
    \frac{\eta^{(n+1)}-\eta^{(n)}}{L_1\Delta t}=&\nabla^2f_1^{(n)}-\nabla^2\eta^{(n+1)}+b_1\nabla^2\phi^{(n+1)}\\
    &-c_1\nabla^4\eta^{(n+1)}
    -\beta\delta\phi^{(n+1)}-\gamma\delta\eta^{(n+1)},\\
    \frac{\phi^{(n+1)}-\phi^{(n)}}{L_2\Delta t}=&\nabla^2f_2^{(n)}-\nabla^2\phi^{(n+1)}+b_1\nabla^2\eta^{(n+1)}\\
    &-c_2\nabla^4\phi^{(n+1)}
    -\alpha\delta\phi^{(n+1)}-\beta\delta\eta^{(n+1)},
\end{aligned}
\right.
\end{equation*}
where $(\eta^{(n)},\phi^{(n)})$ is the current value of $\eta$ and $\phi$, and $(\eta^{(n+1)},\phi^{(n+1)})$ is their next updated value. 
The nonlinear terms are defined by $f_1^{(n)} = f_1\{\eta^{(n)},\phi^{(n)}\}$ and $f_2^{(n)} = f_2\{\eta^{(n)},\phi^{(n)}\}$.

In order to obtain transition pathways connecting different local minima of the phase field model and the corresponding energy barrier of the transitions, 
the string method\cite{weinan2002string,weinan2007simplified}, which is an effective numerical method to obtain minimum energy paths (MEPs)
corresponding to the most probable transition pathways, is applied to our model free energy functional. It is noted that the string method
has been successfully used on particle model to investigate the transition between two opposed membranes to a stalk\cite{muller2012transition} 
and to field-theoretical model to study the pore formation\cite{ting2018metastable}.
One complication of applying the string method is that our phase fields are conserved order parameters, so that the computation of MEP and critical nucleus is subject to the constraint 
$\int_{\Omega} (\eta-\bar{\eta})\rm{d}\mathbf{r} = 0$ and $\int_{\Omega} (\phi-\bar{\phi}) \rm{d}\mathbf{r} = 0$. This constraint could be treated 
using constrained string method\cite{du2009constrained}. 
It is noted that several equivalent formulations such as the Lagrange multiplier method or the augmented Lagrange multiplier method 
could also be used to obtain MEPs. Since we have used Cahn-Hillard equations to find the stable and metastable states, 
it is effective and natural to apply Cahn-Hilliard type dynamics rather than the Allen-Cahn dynamics with additional Lagrange multiplier. 
We also notice that this method has been proposed in \cite{li2017finding}.

In the numerical implementation of the string method, the initial string is constructed using a linear interpolation parameterization connecting the two minima of the system. 
Specifically, the two ends are specified by the two minima, which are then used to generate the intermediate nodes. 
Let $\{\eta_i^n,\phi_i^n\}, i=0,1,...,N$, denotes the i-th node on the string at n-th time step, the steps to obtain the MEP are as follows:\\
\begin{itemize}
\item Step 1 : evolution of the string according to the Cahn-Hilliard equations.
\begin{equation*}
\left\{
\begin{aligned}
    \frac{\eta_i^{*}-\eta_i^{(n)}}{L_1\Delta t}=&\nabla^2 {f_1}_i^{(n)}-\nabla^2 \eta_i^{*}+b_1\nabla^2 \phi_i^{*}-c_1\nabla^4 \eta_i^{*}\\
    &-\beta\delta \phi_i^{*}-\gamma\delta \eta_i^{*},\\
    \frac{\phi_i^{*}-\phi_i^{(n)}}{L_2\Delta t}=&\nabla^2{f_2}_i^{(n)}-\nabla^2\phi_i^{*}+b_1\nabla^2\eta_i^{*}-c_2\nabla^4\phi_i^{*}\\
    &-\alpha\delta\phi_i^{*}-\beta\delta\eta_i^{*},
\end{aligned}
\right.
\end{equation*}
\item Step 2 : string reparametrization with a linear interpolation which conserves concentration parameter.
The arc lengths corresponding to the current images are given by,
\begin{equation*}
    \begin{aligned}
        s_0 &= 0,\\ 
        s_i = s_{i-1}+|\eta_i^*-\eta_{i-1}^*|_2&+|\phi_i^*-\phi_{i-1}^*|_2, i = 1,2,...,N.
    \end{aligned}
\end{equation*}
The mesh is then obtained by normalizing $\{ \alpha_i^* \} = \frac{s_i}{s_N}$.
The new string with equal arc length spacing, $\{\frac{i}{N}, \eta_i^{n+1},\phi_i^{n+1}\}$, is obtained by using linear interpolation from the data $\{\alpha_i^*,\eta_i^*,\phi_i^*\}$.\\
\item Finally, the string of morphologies converges to the MEP. 
\end{itemize}
For a given steady-state string corresponding to a MEP connecting two minima, the node with highest free energy represents the transition state.
This state is an unstable saddle point and the energy barrier between the two minima can be calculated by the energy difference between the initial reactant and the transition state. 

\section{Numerical Results: Equilibrium Structures and Fusion Pathways}

\begin{figure}[htbp]
    \begin{center}
        \includegraphics[width=\columnwidth]{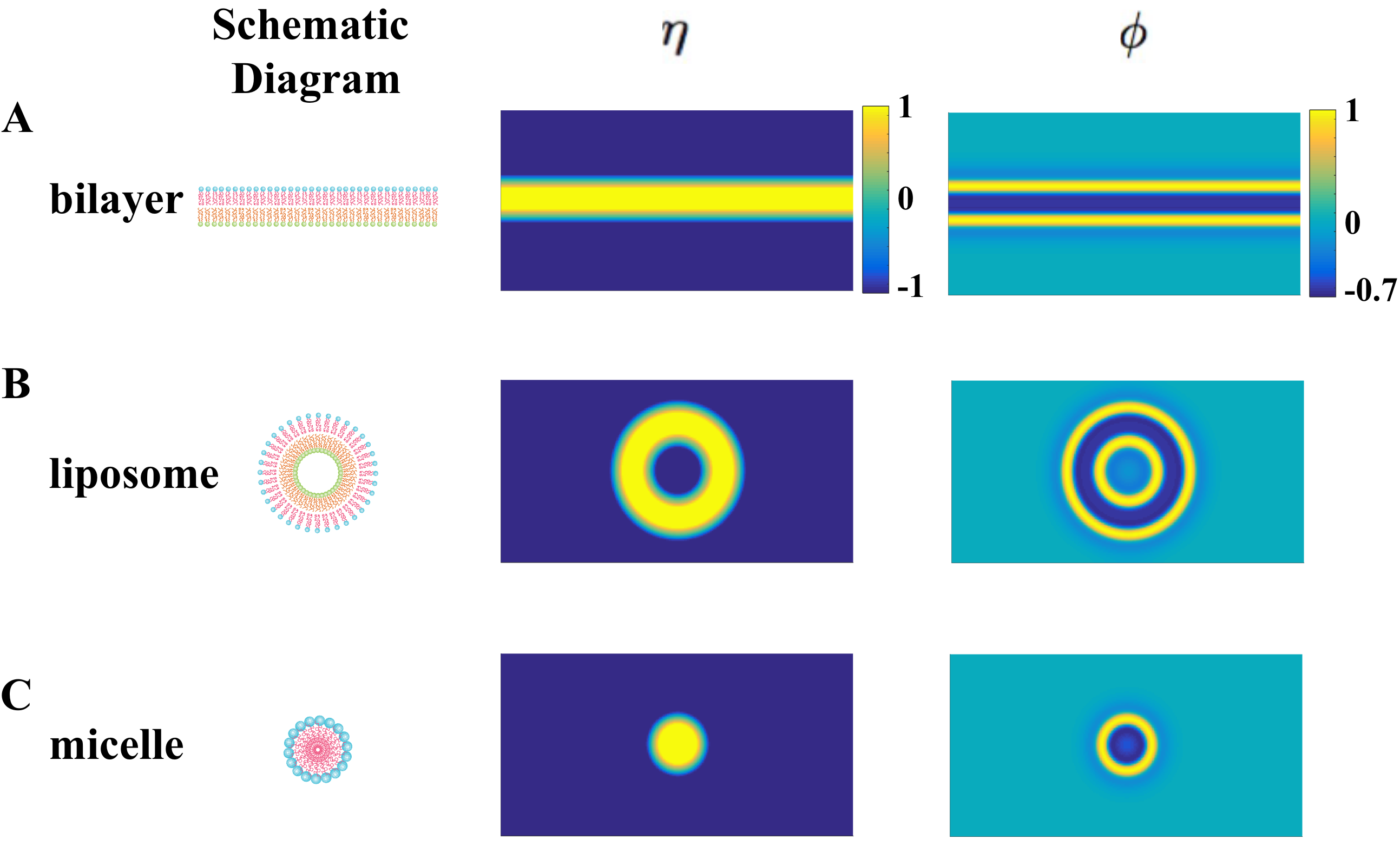}
        \caption{Three equilibrium structures: bilayer(A) at $\bar{\eta} = -0.625$, $\bar{\phi} = 0.005$, liposome(B) at $\bar{\eta} = -0.7$, $\bar{\phi} = 0.004$ and micelle(C) at $\bar{\eta} = -0.94$, $\bar{\phi} = 0.0008$. 
        The first column shows the schematics of the structures, the second and third column show the spatial distributions of the $\eta$ and $\phi$ fields. 
        In the plot of $\eta$, solvent molecules are represented by dark blue and amphiphilic molecules are indicated by yellow. 
        In the plot of $\phi$, solvent molecules are presented by green, hydrophilic head by yellow and hydrophobic tail by dark blue. }
    \label{equilibria}
    \end{center}
\end{figure}

Under different conditions (such as the electrolyte or lipid concentration, pH, and temperature), lipids can associate into various structures in aqueous solutions. 
As the conditions change, the self-assembly structure could transform from one to another\cite{israelachvili2011intermolecular}.
In the phase field model, the parameters controlling the equilibrium structures are the average concentration of the amphiphiles, $\bar{\eta}$.
At different values of $\bar{\eta}$, different equilibrium structures were obtained. 
 Specifically, a planar bilayer structure (Fig. \ref{equilibria}A) is obtained as a stable state of the free energy when $\bar{\eta} = -0.625$. 
Decreasing the value of $\bar{\eta}$ results in the formation of a liposome (Fig. \ref{equilibria}B), {\it i.e.}, a spherical vesicle with one bilayer as the steady state.
Further decreasing $\bar{\eta}$ leads to the formation of a micelle (Fig. \ref{equilibria}C) as the minimizer of the free energy. 
In the micelle, the hydrophobic tail are packed in the core surrounded by the hydrophilic heads.
The fact that these different structures have been obtained as solutions minimizing the free energy of our phase field model demonstrates that our model system is capable
of describing the equilibrium structures of amphiphilic solutions. 

\begin{figure}[htbp]
        \includegraphics[width=1\columnwidth]{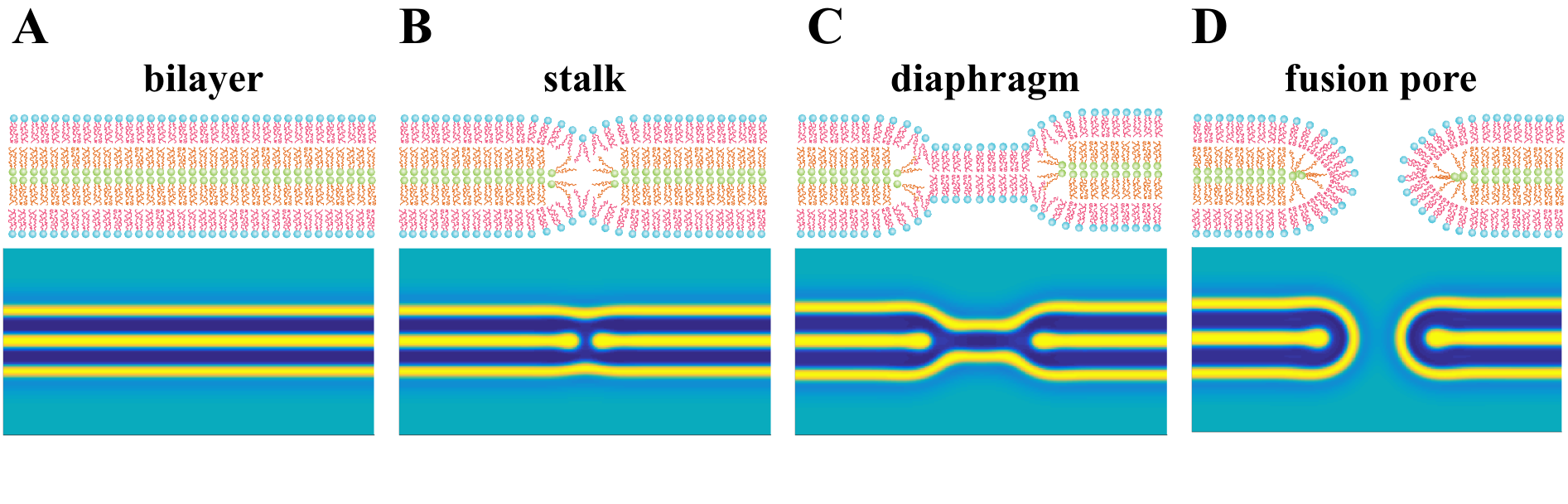}
    \caption{Four steady states with no solvent between membranes correspond to bilayers (A), stalk (B), diaphragm (C) and fusion pore (D), respectively. 
    Pictures in the first row are schematics of lipid molecules where green head with two orange tails indicates \textit{cis} monolayer molecules 
    and blue head with two pink tails indicates \textit{trans} monolayer molecules. Pictures of $\phi$ with $\bar{\phi} =0.01, \bar{\eta} = -0.4$ are in the second row.}
    \label{classical_steady}
\end{figure}

The first step of membrane fusion is to bring them into close contact. It is believed that certain fusion peptides embedded 
in the membranes play an important role in this initial step~\cite{chernomordik2003protein,cai2017liquid}.
In the current work, we assume that this initial step has already been accomplished such that two opposed bilayers have been brought into close contact.
For a model system composed of two opposite membranes in close proximity, the phase field model admits four steady-state solutions shown in Fig. \ref{classical_steady} 
corresponding to (1) opposite bilayers (Fig. \ref{classical_steady}A), (2) stalk (Fig. \ref{classical_steady}B), (3) diaphragm (Fig. \ref{classical_steady}C) and (4) fusion pore (Fig. \ref{classical_steady}D). 
For the case of two opposite bilayers in close contact (Fig. \ref{classical_steady}A), there are no solvent molecules between the membranes.
The hydrophilic heads in the opposed \textit{cis} leaves are in close contact, forming a sandwich-like structure with one hydrophilic layer between two hydrophobic layers. 
The stalk shown in Fig. \ref{classical_steady}B is an hourglass-like structure, in which the hydrophilic and hydrophobic layers are deformed resulting in a hydrophobic bridge. 
The structural change from two opposite bilayers (Fig. \ref{classical_steady}A) to a stalk (Fig. \ref{classical_steady}B) is localized in the central region of the system. 
It is found that the stalk has the highest free energy among the four steady-state structures which means it is less stable than the other structures. 
The structure of the diaphragm (Fig. \ref{classical_steady}C) could be viewed as a stretched stalk with only one bilayer in the middle of the morphology.
Finally, the fusion pore (Fig. \ref{classical_steady}D) is a structure with a pore formed resulting in the fusion of membranes. 
When the fusion pore is formed, the two vesicles become connected through the pore. 
It is found that the fusion pore has the lowest free energy among four steady-state structures and is the most stable structures.

\begin{figure*}
        \centering
        \includegraphics[width=1.8\columnwidth]{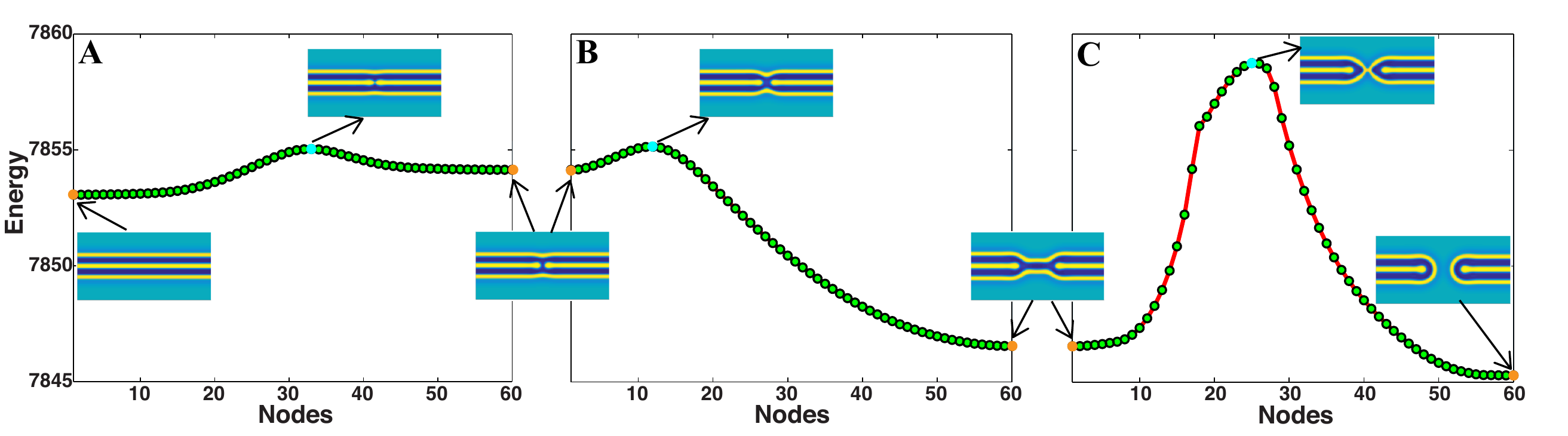}
        \caption{Transition pathway of classical path with no solvent between two bilayers. (A-C) depict paths from two opposed bilayer to stalk, from stalk to diaphragm and from diaphragm to fusion pore, respectively. In each picture the first and last node on the string are steady states and the saddle points in (A-C) are the 33th, 12th and 25th node of string, respectively.} 
    \label{classical_MEP}
\end{figure*}
For the case of two bilayers in close contact, a MEP connecting these four steady states (bilayers, stalk, diaphragm and fusion pore) has been obtained by applying the string method to the phase field model. 
This MEP shown in Fig. \ref{classical_MEP} corresponds to the complete classical transition pathway and it is the most probable transition path from two opposite bilayer membranes to a fusion pore. 
Along the transition pathway from bilayers to stalk (Fig. \ref{classical_MEP}A), the membrane becomes thinner in the middle of the system. 
A critical state, corresponding to the maximum of the energy along the MEP, is reached when the two initially separated hydrophilic regions 
touch each other (Fig. \ref{classical_MEP}A) initiating a hydrophobic bridge connecting the top and bottom hydrophobic layers. 
After the critical state, the energy decreases along the MEP when the hydrophobic bridge widens until a stalk structure is formed. 
The transition from two opposite bilayers to the stalk has been studied by a number of researchers~\cite{muller2003new, kuzmin2001quantitative}.
One noticeable observation is that there is a relatively large energy barrier to form the critical state along this MEP. Therefore the transition from
two opposite bilayers to a stalk needs to over come a large energy barrier, thus the process would not occur spontaneously in general.
In order to overcome the energy barrier, fusion proteins such as SNARE proteins, synaptotagmin and influenza hemagglutinin are needed, 
which could be inserted into a membrane to induce local bending~\cite{martens2008mechanisms, campelo2008hydrophobic, kozlov1998mechanism}. 
Along the transition pathway (Fig. \ref{classical_MEP}B) from the stalk to diaphragm, the hydrophobic bridge expands horizontally and shrinks vertically,
resulting in the diaphragm composed of a bilayer connecting two sandwich-structures. 
The transition pathway from the diaphragm to fusion pore is shown in Fig. \ref{classical_MEP}C. 
It is interesting to note that the pore is not formed by rupturing the bilayer in the middle of the diaphragm. 
Rather, it is observed that the diaphragm contracts horizontally until a critical state, in which the two bilayers are connected by a nipple-like structure, is reached.
After this critical state, a pore is formed connecting two interior sides of the bilayers.
The energy of the system decreases as the pore becomes larger until a steady state corresponding to the fusion pore is reached.

\begin{figure}[htbp]
        \includegraphics[width=1\columnwidth]{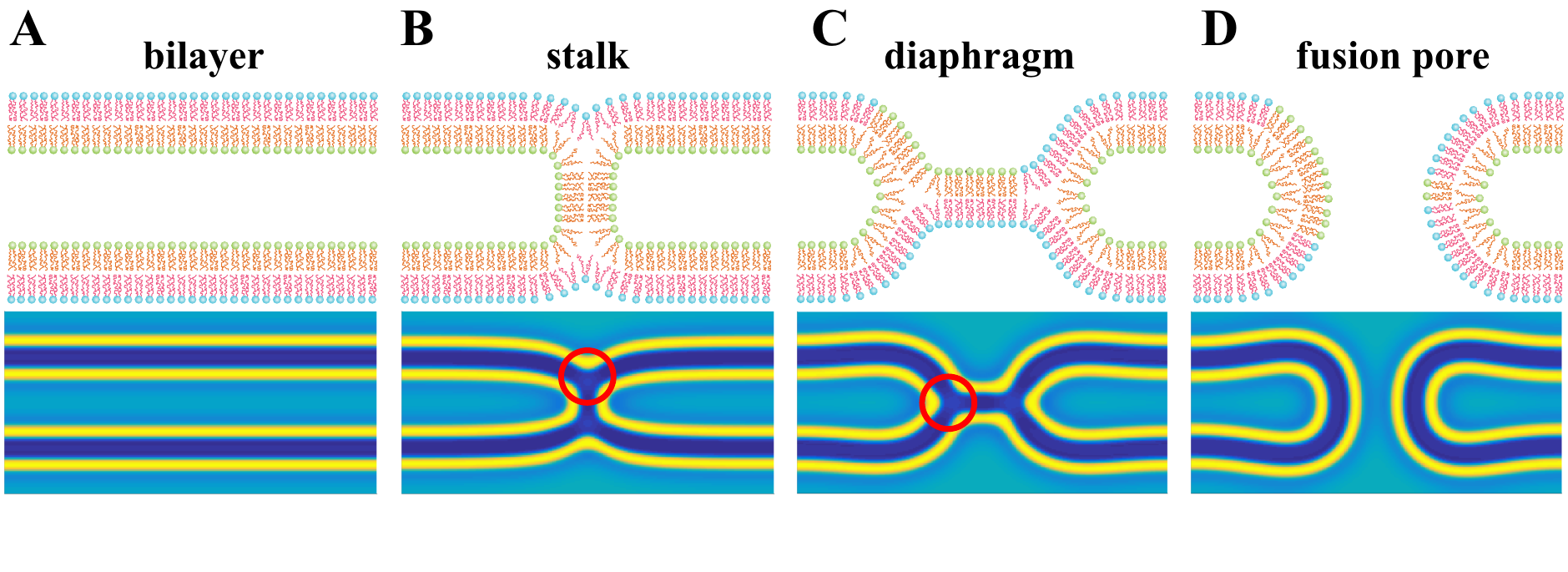}
    \caption{Four steady states correspond to bilayer (A), stalk (B), diaphragm (C) and fusion pore (D), respectively. Pictures in the first row are detail displacement of lipid molecules where green head with two orange tails indicates \textit{cis} monolayer molecules and blue head with two pink tails indicates \textit{trans} monolayer molecules. Pictures of $\phi$ are with $\bar{\eta} = -0.25$ in the second row. Red circles indicate the place of dimples.}
    \label{steady}
\end{figure}

In reality, the two opposite bilayers may not be in close contact. In this case a small gap composed of solvent molecules exists between the two bilayers.
When there is a gap between the two opposed bilayers, the phase field model still possesses four steady-state solutions corresponding to the state of 
bilayers (Fig. \ref{steady}A), stalk (Fig. \ref{steady}B), diaphragm (Fig. \ref{steady}C) and fusion pore (Fig. \ref{steady}D). 
These structures are similar to the corresponding steady states of the case when two opposite bilayers are in close contact with the exception that
there is a gap between the two bilayers. 
{Computationally, the free energy of the phase field model is minimized when the distance between the two bilayers is equal to a fixed value, which depends on the parameters. One can observe that, at the left and right boundaries, the distance between the two bilayers almost stay unchanged during the whole fusion process.}
Experimentally, a gap between two opposite bilayers could be maintained via a number of mechanisms. 
In the experiments reported in~\cite{tamm2003membrane}, a fixed distance due to the balance of Van der Waals attraction and hydration repulsion is observed. The distance between electrically neutral bilayers ranges from 1.2 to 2.2nm~\cite{rand1989hydration}, which is about one or two membranes thickness. 
From the computational results shown in Fig. \ref{steady}, it is noted that the membranes in stalk (Fig. \ref{steady}B), diaphragm (Fig. \ref{steady}C) and fusion pore (Fig. \ref{steady}D) 
are heavily curved when compared to the structures in Fig. \ref{classical_steady}. The large bending of the bilayers is caused by the distance between the two membranes. 
In the stalk (Fig. \ref{steady}B) and diaphragm (Fig. \ref{steady}C) configurations, as the membranes were dented, differently curved monolayers contact each other. 
However, the hydrophobic tails could not be compactly arranged where hydrophobic interstices are formed as indicated by the red circles in Fig. \ref{steady}.

\begin{figure}[htbp]
        \includegraphics[width=1\columnwidth]{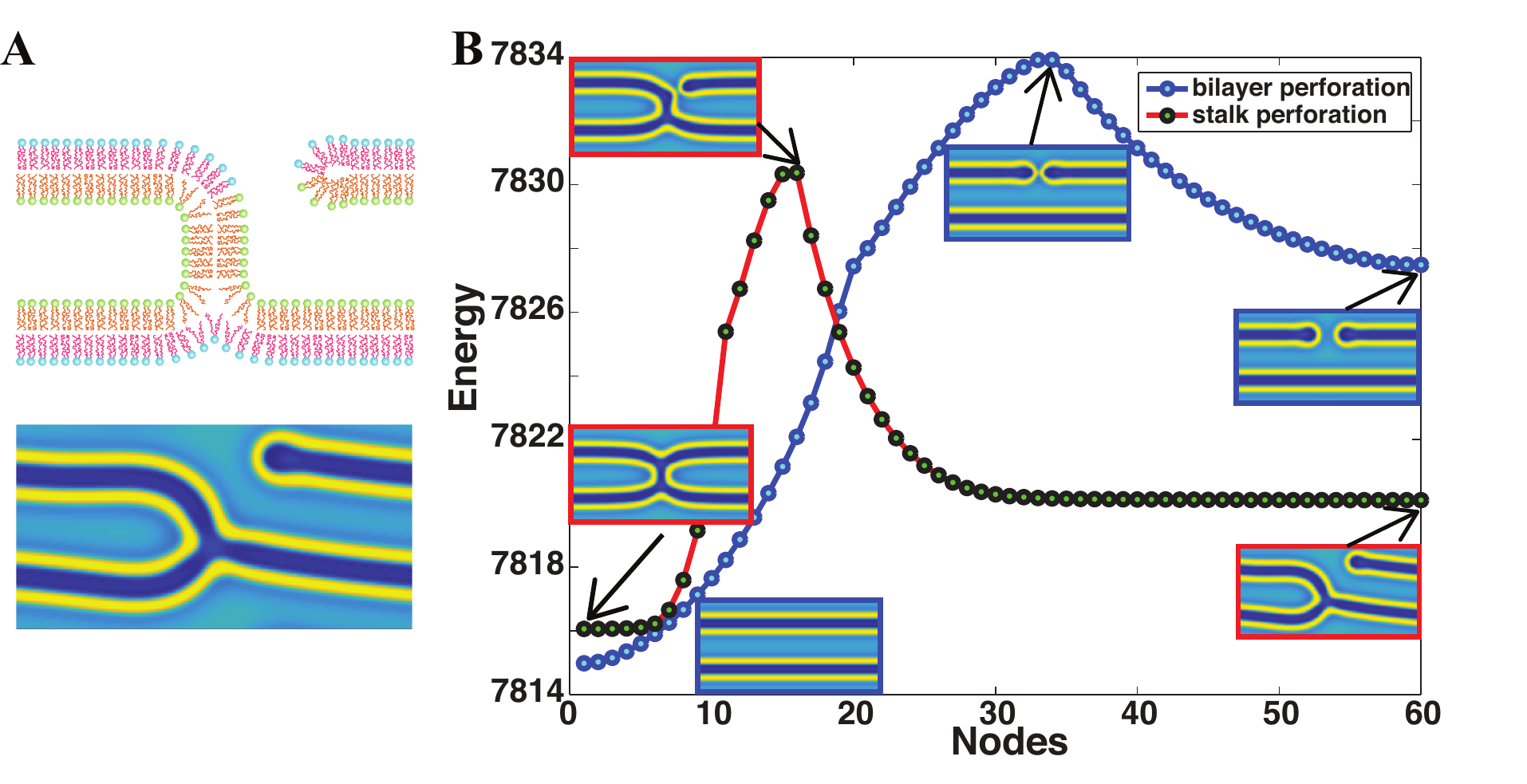}
    \caption{(A) Stalk-hole configuration. The upper image is the sketch of stalk-hole complex. Green and blue balls represent hydrophilic heads of \textit{cis} and \textit{trans} leaves, respectively, orange and pink curves represent hydrophobic tails of \textit{cis} and \textit{trans} leaves, respectively. 
The lower image is simulation result of $\phi$
with $\bar{\eta} = -0.25$. 
(B) The red line with green bead and blue line and blue bead indicates MEP of stalk perforation and bilayer perforation, respectively. 
    } 
    \label{stalk-hole_compound}
\end{figure}
\begin{figure*}
        \centering
        \includegraphics[width=2\columnwidth]{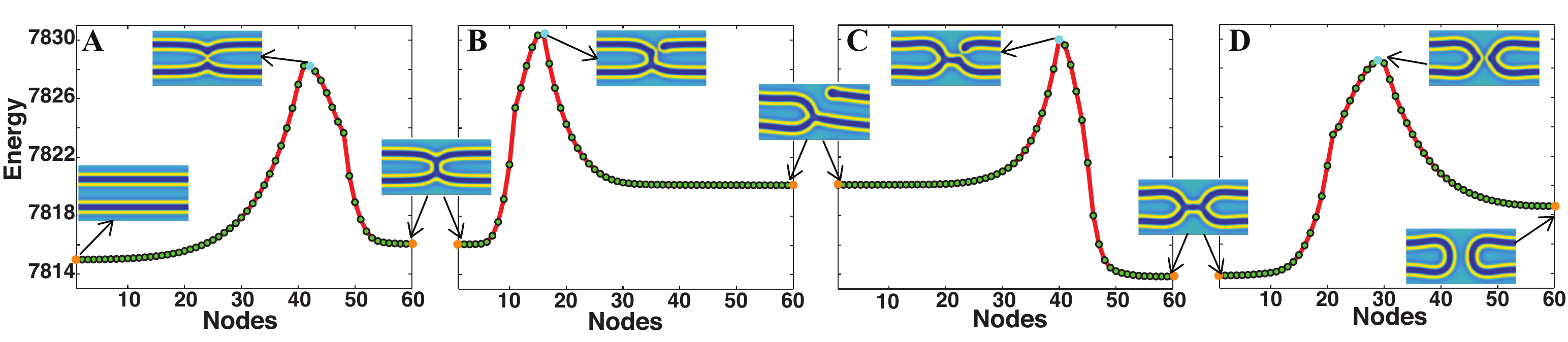}
    \caption{
        (A-D) depict transition pathways of leaky path from two opposed bilayer to stalk, from stalk to stalk-hole, from stalk-hole to diaphragm and from diaphragm to fusion pore, respectively. In each picture the first and last node on the string are steady states and the saddle point in (A-D) is the 42th, 16th, 40th and 29th node of string, respectively.
    } 
    \label{leaky_MEP}
\end{figure*}

The existence of hydrophobic interstices will hinder the transition between the stalk and diaphragm configurations.
We tried to obtain a rotationally symmetric MEP for the case of two opposite bilayers with a gap between them and discovered that 
there exists an unstable high-energy intermediate state between the stalk and diaphragm in which the two interstices merge into a relatively large one. 
When the condition of rotational symmetry is relaxed, we obtained a steady state corresponding to a new stalk-hole complex composed of a hole near stalk (Fig. \ref{stalk-hole_compound}A).
Previous simulations~\cite{muller2003new} have shown that the formation of holes is correlated in time and space with the formation of stalks and the presence of a stalk encourages a hole to form nearby. 
The existence of a stalk-hole complex as a steady state along the MEP is consistent with these simulations.
It is informative to compare the energy barrier of bilayer perforation MEP with stalk perforation MEP (Fig. \ref{stalk-hole_compound}B). 
The energy of the stalk-hole steady state is lower than bilayer-hole steady state, indicating that a hole with a stalk nearby is more stable. 
The energy barrier of stalk perforation MEP is only $76\%$ of that of bilayer perforation MEP, i.e., the stalk structure facilitates the formation of hole. 
However, the stalk-hole complex configuration still has higher energy than steady states in Fig. \ref{steady}.

When the condition of rotational symmetry is not maintained, a different MEP going through the stalk-hole configuration has been obtained.
The complete leaky transition pathway shown in Fig. \ref{leaky_MEP}A-D is the MEP connecting the steady states corresponding to the
bilayers, stalk, stalk-hole complex, diaphragm and fusion pore.
In Fig. \ref{leaky_MEP}A bulges emerge from the middle of two initially flat bilayers. The shape of the inner \textit{cis} monolayers and the outer \textit{trans} monolayers 
deform simultaneously to maintain a constant thickness of membranes. The \textit{cis} monolayers touch first with two stretched tips and the hydrophilic 
core which separates the hydrophobic passage is broken through. When the passage is as wide as one bilayer, a stalk is formed. 
In Fig. \ref{leaky_MEP}B, a hole emerges near the stalk in the upper membrane and the lipid molecules surrounding it are rearranged 
so that the hydrophobic tails are separated from solvent by hydrophilic heads. With a hole, the inside and outside solvent is connected which means the existence of a leakage. 
In Fig. \ref{leaky_MEP}C, the upper right membrane is connected to lower right membrane. After adjustment, diaphragm is formed. 
It is noteworthy that the \textit{cis} leaf and \textit{trans} leaf mix together. From diaphragm to fusion pore (Fig. \ref{leaky_MEP}D), 
the connection between left and right part changes from one bilayer to a nipple-structure, which is the critical state. 
After the nipple-structure disappears, the energy is decreasing as the distance between left and right part membrane is larger, until the fusion pore is reached. 
The energy barrier of the MEP from bilayer to stalk without solvent between two bilayers (Fig. \ref{classical_MEP}A) is greatly lower than that with solvent (Fig. \ref{leaky_MEP}A), 
while the energy barrier of MEP from diaphragm to fusion pore (Fig. \ref{classical_MEP}C) is almost the same as the energy barrier in Fig. \ref{leaky_MEP}D. 
The energy barrier from stalk to diaphragm in the classical transition path (Fig. \ref{classical_MEP}B) compared with that from stalk to diaphragm via stalk-hole complex (Fig. \ref{leaky_MEP}B-C) is much lower.

\section{Conclusion and Discussion}

In this work, we propose a molecular-informed phase field model for the study of structural transitions of self-assembled bilayers from amphiphilic molecules dissolved in 
hydrophilic solvents. We emphasize that, in order to describe the details of the bilayer structures, the phase field model has to be developed using information of the
molecular properties of the system. In the case of amphiphilic molecules dissolved in solvents, the fact that an amphiphilic molecule is composed of hydrophilic and 
hydrophobic parts requires a model system which is capable of describing the microscopic phase separation of the hydrophilic heads and hydrophobic tails. 
We introduce two phase fields to describe macroscopic solvent/amphiphile separation and microscopic amphiphilic head/tail separation, respectively. 
The energy functional of the phase field model, or the Landau free energy of the system, is developed by
extending the theory of block copolymers proposed by Ohta and Kawasaki. The molecular information is encoded in the phase field energy functional.
In particular, the fact that the hydrophilic heads and hydrophobic tails are connected to form one molecule is described by the long-range term of the energy functional.
This phase field model is capable of predicting the formation of self-assembled bilayers. A combination of the phase field model and the string method is used
to obtain minimum energy paths connecting two opposite bilayers with a fusion pore. 
In the case of membrane fusion, we obtained two complete transition pathways from two opposed bilayers to fusion pore. 
When the two opposite bilayers are in close contact, a MEP corresponding to the classical transition pathway is obtained. 
It is noted that the classical transition path has been proposed previously. The results obtained in the current work are in agreement with available results.
When the two opposite bilayers are separated by a small gap, a different MEP, corresponding to a leaky transition pathway is obtained. 
This MEP is characterized by the presence of a stalk-hole complex as an intermediate state. 
The structural features and information about phase transition processes, such as transition state and energy barrier, are contained in the results naturally. 

Numerical study of the proposed phase field model can be naturally extended to 3D membrane fusion computation. 
More features of stalk-hole complex configuration would be revealed in 3D calculations and the leaky transition pathway through 
stalk-hole complex configuration may also be different. More MEPs from two opposed bilayers to a fusion pore should be expected. 
In the current work, we focus on the model development and validation, and we will leave the extended 3D calculations to future study. 

{\bf Acknowledgment}
This work was funded by the National Natural Science Foundation of China No. 11622102, 11861130351, 11421110001 and by the Natural Science and Engineering Research Council (NSERC) of Canada.

\bibliographystyle{unsrt}
\bibliography{MF_arxiv}
\end{document}